\documentclass{trbunofficial-ad}

\usepackage{graphicx}
\usepackage{amsmath}
\usepackage{amssymb}
\usepackage{booktabs}
\usepackage{siunitx}
\usepackage{multirow}
\usepackage[hidelinks]{hyperref}
\usepackage{dsfont}

\graphicspath{{../}{./}}

\makeatletter
\long\def\@makecaption#1#2{%
  \vskip\abovecaptionskip
  {\raggedright\bfseries\boldmath #1\;#2\par}%
  \vskip\belowcaptionskip}
\makeatother

\usepackage[backend=biber, style=authoryear, sorting=nyt, natbib=true,
            giveninits=false, doi=true, url=false, isbn=false, eprint=false]{biblatex}
\usepackage{zref-totpages}
\DeclareFieldFormat{title}{\mkbibquote{#1}}
\DeclareFieldFormat[article]{title}{\mkbibquote{#1}}
\AtBeginBibliography{}

\AuthorHeaders{R. Bello, J. Mwakalonge, V. Vaidyan, V. Bendigeri, A. Dontoh, G. Comert, M. Mkufya}
\title{Recovering Hidden Wildlife--Vehicle Collisions from Existing Crash Data:
A Cross-State Framework for Detection, Prevalence Estimation, and
Under-Reporting Audits}

\author{%
  \textbf{Rasheed Bello*}\\
  South Carolina State University\\
  Orangeburg, SC 29117, USA\\
  ORCID: https://orcid.org/0009-0007-7918-1703\\
  Email: rbello@scsu.edu\\
  \hfill\break
  \textbf{Judith Mwakalonge}\\
  South Carolina State University\\
  Orangeburg, SC 29117, USA\\
  Email: jmwakalo@scsu.edu\\
  \hfill\break
  \textbf{Varghese Vaidyan}\\
  Dakota State University\\
  Madison, SD 57042, USA\\
  Email:varghese.vaidyan@dsu.edu\\
  \hfill\break
  \textbf{Vijay Bendigeri}\\
  Independent Researcher\\
  Sacramento, CA, USA\\
  Email: vijaybendigeri@gmail.com\\
  \hfill\break
    \textbf{Anthony Dontoh}\\
  South Carolina State University\\
  Orangeburg, SC 29117, USA\\
  Email: adontoh@scsu.edu\\
  \hfill\break
 \textbf{Gurcan Comert}\\
  North Carolina A\&T State University\\
  Greensboro, NC 27411, USA\\
  Email: gcomert@ncat.edu\\
  \hfill\break
    \textbf{Miriam Mkufya}\\
  South Carolina State University\\
  Orangeburg, SC 29117, USA\\
  Email: mmkufya@scsu.edu\\
\vspace{0.4em}
\hfill
}
\DeclareSIUnit{\mile}{mi}

\begin{document}

{\fontsize{12}{14.4}\selectfont
\maketitle}
\thispagestyle{empty}
\newpage

\pagestyle{main}
\section*{Abstract}

\noindent\textbf{Objectives}\hspace{0.6em}
Wildlife--vehicle collisions (WVCs) impose a large and poorly measured toll on
United States highways, yet most state crash databases never record which
crashes involved an animal, so agencies cannot say how many they miss. This
paper asks whether the crash records agencies already hold can be used to
identify WVCs at the record level and to estimate how many go uncoded.\\

\noindent\textbf{Methods}\hspace{0.6em}
We assembled 4.9 million crash records from Colorado, Utah,
Iowa, and Illinois, spanning 2007 to 2026. Each state's schema was passed through a label-leakage screen that removes any field encoding animal involvement, and separately harmonized to a shared vocabulary for cross-state analysis. Labeled wildlife vehicle crashes are treated as ground-truth, with which  Calibrated gradient-boosted classifiers
were trained per state and read with Shapley additive explanations. Prevalence
was then estimated directly from classifier scores using Adjusted
Classify-and-Count and Maximum-Likelihood Prevalence Estimation, validated
in-domain, across states, and against injected reporting gaps.\\

\noindent\textbf{Findings}\hspace{0.6em}
Detection quality is governed by how completely a crash is recorded, not by how
rare WVCs are: at equivalent prevalence near four percent, feature-rich Utah
reaches an average precision of 0.88 against 0.64 for feature-poor Colorado,
and Utah exceeds Iowa despite Iowa's threefold higher prevalence. A within-state
richness curve with bootstrap confidence intervals confirms the effect with
prevalence held fixed. Models rank records almost perfectly everywhere, but
their calibration does not transfer, and prevalence estimates recover the true
rate to within 0.001 in-domain.\\

\noindent\textbf{Novelty}\hspace{0.6em}
This work reframes WVC surveillance from predicting where collisions cluster to
detecting which records are collisions and estimating how many are missing,
using only data agencies already collect.\\

\noindent\textbf{Practical Applications}\hspace{0.6em}
Agencies can retrospectively flag unlogged WVCs in legacy files, quantify their
true collision burden for mitigation funding, and learn which crash-report
fields most improve detection before revising a reporting manual.

\vfill
\noindent\textit{Keywords}: wildlife--vehicle collisions, crash records, under-reporting,
quantification learning, gradient boosting, data quality

\newpage
\pagestyle{main}
\setcounter{page}{1}

\section{INTRODUCTION}

Collisions between vehicles and large wild animals are among the most common
crash types on rural highways yet among the least visible in the data that
transportation agencies use to set priorities \citep{bell2024, jorgensen2025}. Wildlife-Vehicle Crashes (WVC) cost the United States above eight billion dollars, alongside tens of
thousands of injuries and close to two hundred deaths \citep{bell2024}. Earlier
national accounting put large-mammal collisions at one to two million per year
\citep{huijser2009}. Texas alone recorded 51,522 animal-related crashes between
2010 and 2016, costing motorists roughly \$1.3 billion annually
\citep{wilkins2019}. These figures share a fundamental limitation. Every one of
them is built from crashes that were reported, coded, and retained, and the
literature is unanimous that this is a fraction of what occurs. Maintenance
crews recover five to ten carcasses for every collision that reaches a police
report \citep{wilkins2019}, and Montana's transportation department counted
60,858 carcasses over a decade against 23,718 collisions logged by highway
patrol over the same period \citep{bell2024}. Despite the significant disparities, there has been no unified method to  estimate the cost of unreported WVCs.

This needed capability to address this gap is hinged on state crash reporting rules, which establishes what is recorded into  crash report, and how. When North Dakota narrowed its reporting
requirement to injury and fatality crashes in 2013, annual reported WVCs fell
from 3,488 to 133, and seasonal effects that had been statistically detectable
before the change became invisible afterwards purely through loss of sample
size \citep{jorgensen2025}. The same administrative dynamics that decides
whether a crash is recorded also decides how richly it is described, and both
decisions propagate into every downstream analysis an agency performs.

Machine learning (ML) has been applied to this problem from a different perspective. A decade of
work has treated WVCs as a spatial phenomenon and asked where collisions
concentrate, fitting risk surfaces to road segments or grid cells
\citep{gunson2011, bell2024, pagany2022}. These ML applications are crucial when siting a wildlife crossing structures but they do not answer the question: which of these
records are wildlife collisions, and how many did we fail to record? A risk map
cannot answer either. It assigns a severity score to a place, not a label to a record,
and it is typically fitted against synthetic non-collision points rather than
against the real crash records in the agency's database.

\begin{figure}[!ht]
  \includegraphics[width=\linewidth]{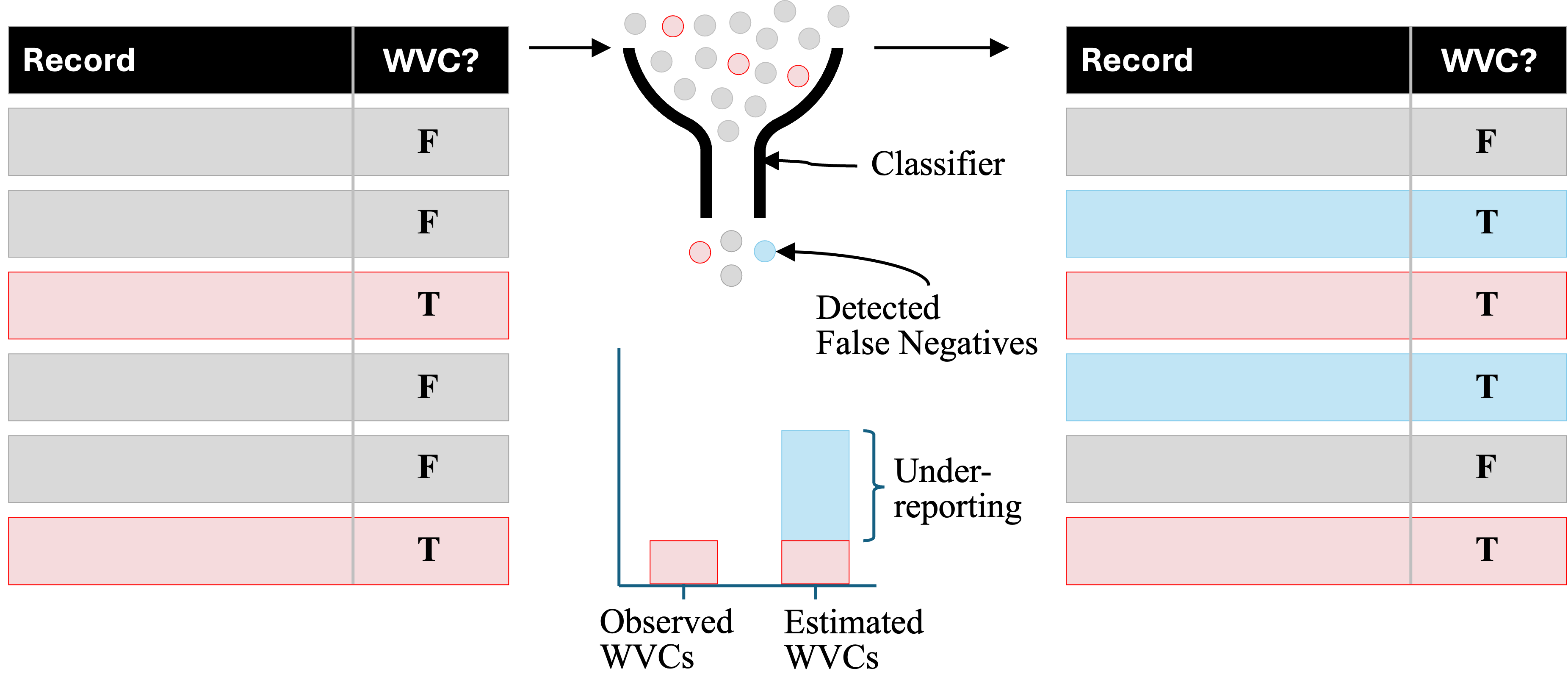}
  \caption{Research Framing: Record-level detection and
  prevalence estimation. \textit{T = True, F=False}}
  \label{fig:framing}
\end{figure}

We address this crucial gap accordingly. Treating each crash record as the unit of
analysis turns WVC surveillance into two coupled problems: detecting which
records describe animal involvement using only fields that do not encode the
answer, and estimating what fraction of records are WVCs when the coded label is
itself unreliable. The second problem has an established solution outside
transportation. Quantification learning estimates class prevalence in a target
sample directly from classifier scores, and does so reliably even when
record-level classification is mediocre and the positive class is rare
\citep{forman2008, gonzalez2018}. Applied to crash data, it converts a
classifier into an auditing instrument: the difference between the estimated
prevalence and the agency's own coded rate is a direct measure of
under-reporting. Figure~\ref{fig:framing} places the three questions in
sequence.

This paper makes five contributions. First, it establishes that record-level WVC
detection is feasible from routinely collected structured fields alone, across
four states and 4.9 million crashes, without any field that names an animal.
Second, it shows that the ceiling on detection is set by how rich a crash
is described rather than by how prevalent WVCs are, and defends that claim with a
within-state analysis that holds prevalence fixed while varying record
richness. Third, it introduces a reusable label-leakage screen that both
prevents circular models and doubles as a reporting-practice audit, revealing
that one state encodes animal involvement redundantly across ordinary
descriptor fields. Fourth, it decomposes cross-state transfer, showing that
record ranking survives transfer while calibration and decision thresholds do
not, which bounds what a borrowed model can legitimately be used for. Fifth, it
recovers WVC prevalence without labels and converts the result into an
under-reporting estimate that agencies can compute on their own files.

\section{RELATED WORK}

\subsection{A Burden Measured by Its Absence}

The WVC under-reporting literature has circled the same quantity for a decade
without estimating it. \citet{snow2015} approached it as a robustness question,
deleting reported deer and moose collisions at increasing rates and watching
logistic-regression coefficients and predictive AUC degrade. Their reassuring
answer, that inference survives until roughly seventy percent of records are
missing, rests on WVCs clustering in low-variance hotspots, and it licenses the
continued use of imperfect databases. \citet{jorgensen2025} complicate that
optimism from the opposite side. Where Snow and colleagues simulated random
loss, they document a policy-driven loss that pushed North Dakota's reported
counts far below any tolerable threshold, and they show the damage is done not
by bias but by sample-size collapse. Read together, the two papers bracket the
risk: under-reporting is survivable until a reporting rule changes, at which
point it is not.

Neither study attempts recovery, and the field's two strongest empirical
anchors stop at the same boundary. \citet{wilkins2019} quantify the Texas burden
in dollars and document the carcass-to-report ratio that defines the missing
mass, but model county-level counts by ordinary least squares and take the
animal-related label as given. \citet{zou2019} come closest to the target,
coupling a negative-binomial collision-count model to a latent reporting process
through a Gaussian copula and showing that ignoring under-reporting biases
covariate effects. Their approach is parametric, operates on segment counts
rather than records, and requires carcass removals as a second measurement of
the same process. Where carcass data do not exist, and in most states they do
not, the method has nothing to condition on. \citet{ascensao2019} widen the
frame usefully by arguing that a zero in a mortality record can mean
extirpation or a barrier effect rather than low risk, though as a perspective
piece it supplies motivation rather than machinery.

\subsection{Machine Learning That Answers ``Where''}

The predictive literature inherited its question from the review that organized
it. \citet{gunson2011} synthesized twenty-four generalized linear models of WVC
location, catalogued the predictors that recur, warned that many published
effects are confounded, and called for models whose predictive power is
demonstrated in landscapes other than the one they were fitted in. Subsequent
work answered the call for better estimators more readily than the call for
honest transfer testing.

\citet{bell2024} treat Montana's segment-level carcass counts as an imbalance
problem, using SMOTE to rebalance zero-heavy segments before fitting negative
binomial and random forest models. The framing is instructive because it is the
one our results contradict: rarity is cast as the obstacle, and resampling as
the remedy. \citet{pagany2022} push furthest on generalization, fitting random
forests, neural networks, and support vector machines to roughly 731,000
Bavarian police-reported WVCs at hourly and 25-metre resolution, then testing
transfer both spatially and forward in time, and reporting success on both. That
success deserves qualification. The negative class consists of randomly
generated non-accident points, so the model never distinguishes a wildlife
crash from an ordinary one; the training class ratio is chosen rather than
observed, which sidesteps prevalence entirely; and transfer is tested between
regions of a single country under one reporting regime. None of these studies
screens for fields that encode the outcome, and none reports calibration.

\subsection{Detecting Records That Were Labelled Wrongly}

The methodological analog to our task sits outside the WVC literature, in work
that treats the crash database itself as the object of correction.
\citet{hosseini2023} apply text mining to police narratives to separate genuine
wrong-way driving crashes from records merely flagged as candidates, reaching
roughly 82 percent accuracy and cutting the manual review burden on a rare,
severe crash type, which is structurally the same problem WVCs present.
\citet{bhagat2025} pursue misclassified intersection-related crashes in Iowa,
and report a result that anticipates ours from an independent direction: fusing
narrative text with structured fields cut error rates by 54 percent, which is
the qualitative form of the completeness effect we quantify. Both approaches
depend on free-text narratives, which are frequently unavailable, inconsistently
written, and encumbered by privacy restrictions. Neither estimates prevalence,
and neither tests transfer across agencies.

\subsection{Counting Without Labels}

Estimating the proportion of positives in an unlabelled sample is a task in its
own right, and one that behaves differently from classification.
\citet{forman2008} proved that naive classify-and-count is biased whenever test
prevalence departs from training prevalence, introduced the adjusted estimator
that inverts the classifier's error rates, and demonstrated that prevalence can
be recovered reliably even when per-record accuracy is poor and positives are
rare, which is precisely the regime crash data occupy. \citet{saerens2002} arrive at the
same destination by a different route, re-estimating the target prior through an
expectation-maximization procedure that maximizes the likelihood of the
unlabelled sample and returns the new prevalence as a by-product.
\citet{gonzalez2018} place both within a taxonomy and, importantly for our
design, prescribe evaluating quantifiers by systematically varying test
prevalence rather than reporting a single number.

The failure conditions matter as much as the methods. \citet{esuli2021}
stress-tested the expectation-maximization estimator and found it dependable
only when the classification scheme has few classes and the classifier is
calibrated, degrading and often failing to converge otherwise. That result
anticipates the cross-state behaviour we report.

\subsection{Where This Work Sits}

Four literatures converge here and have not previously been combined. The
under-reporting literature identifies the quantity worth measuring but stops at
documentation or requires a second data source. The WVC machine-learning
literature builds capable models of the wrong question, against synthetic
negatives, without leakage screening or calibration reporting. The crash-record
mining literature demonstrates that mislabelled records can be recovered, but
only through narratives and only within one agency. Quantification learning
supplies the estimator the first literature lacks, together with an explicit
account of when it should not be trusted. This paper joins them on structured
crash fields that every state already collects.

\section{DATA}

We assembled complete crash files from four state departments of
transportation, chosen to span a wide range of both WVC prevalence and record
richness while remaining within a single reporting culture. Table~\ref{tab:states}
summarizes them. The pooled corpus contains 4,912,128 records with valid
coordinates and a non-missing animal label.

Each state marks animal involvement in a single binary field, but not with the
same meaning. Colorado's \texttt{wildlife\_related} and Utah's
\texttt{wild\_animal\_related} refer specifically to wild animals, whereas the
Iowa and Illinois flags are derived from broader animal-crash coding that does
not everywhere separate wild from domestic animals. \citet{wilkins2019} found
domestic animals in 31 percent of Texas animal-related reports, so this
distinction is material and we treat it as a bound on cross-state label
comparability rather than as noise to be ignored. Since every model is trained
and evaluated against the label of its own state, the distinction affects
transfer and cross-state prevalence comparisons but not the per-state results.

Prevalence varies by nearly a factor of four across the four states, from 3.75
percent in Colorado to 13.96 percent in Iowa. Record richness varies
independently: Utah describes a crash with 23 usable fields and Iowa with eight,
despite Iowa's much higher WVC rate. That the two quantities vary independently
is what makes the four states a usable design rather than a convenience sample.

Roadway attributes were attached by a nearest-segment spatial join in the CONUS
Albers equal-area projection (EPSG:5070), matching each crash to the closest
attributed centerline segment within 150~m and transferring annual average daily
traffic, posted speed, through-lane count, an urban--rural indicator, and a
segment sinuosity ratio. Attribute availability differs sharply by state: Utah's
statewide centerline carries traffic attributes only on the state-highway
subset, Utah's through-lane field is effectively empty, and Iowa's route layer
carries no traffic attributes at all and segments too long for sinuosity to
mean anything. Roadway variables therefore enter only the harmonized
cross-state feature set described below, never the per-state models, and Iowa
contributes no roadway attributes.

\begin{table}[!ht]
\centering
\caption{Study states data summary.}
\label{tab:states}
\begin{tabular}{|l|r|r|l|r|r|r|r|}
\hline
\textbf{State} & \textbf{Records} & \textbf{WVCs} & \textbf{Years} & \textbf{Prev.} & \textbf{Cand.} & \textbf{Flag.} & \textbf{Model} \\
\hline
Colorado & 2,028,656 &  76,099 & 2007--2025 &  3.75\% & 10 &  3 & 10 \\
\hline
Utah     &   502,991 &  22,044 & 2018--2026 &  4.38\% & 21 &  1 & 23 \\
\hline
Iowa     &   622,292 &  86,900 & 2015--2026 & 13.96\% & 15 & 10 &  8 \\
\hline
Illinois & 1,758,189 &  93,097 & 2019--2024 &  5.30\% & 19 &  2 & 20 \\
\hline
Total    & 4,912,128 & 278,140 & 2007--2026 &  5.66\% & & & \\
\hline
\end{tabular}

\vspace{0.4em}
\begin{minipage}{\linewidth}
\footnotesize Cand., candidate crash-record fields; Flag., fields removed by the
leakage screen; Model, fields the model uses (leakage survivors plus three
derived temporal fields, month, hour, and day of week, available in every state).
\end{minipage}
\end{table}

\section{METHOD}

\begin{figure}[!ht]
  \includegraphics[width=\linewidth]{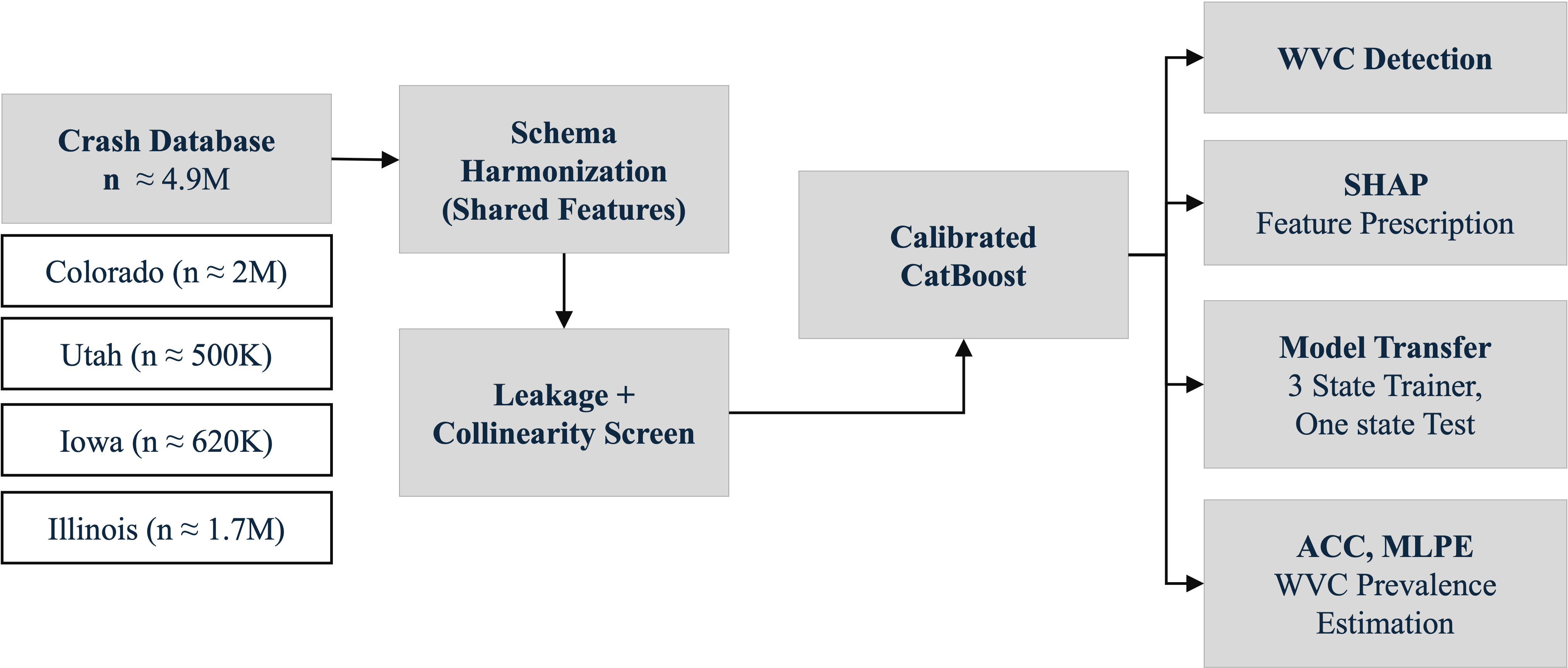}
  \caption{Analysis pipeline.}
  \label{fig:pipeline}
\end{figure}

Figure~\ref{fig:pipeline} summarizes the pipeline. The remainder of this section
develops each stage.

\subsection{Problem Formulation}

Let a crash file be a set of records $\{(\mathbf{x}_i, y_i)\}_{i=1}^{n}$ where
$\mathbf{x}_i$ represents each structured field a responding officer completed and
$y_i \in \{0,1\}$ indicates animal involvement as coded by the agency. Two
quantities are of interest. The first is the posterior $\Pr(y=1 \mid
\mathbf{x})$, which supports record-level detection. The second is the marginal
$\pi = \Pr(y=1)$, the prevalence, which supports burden estimation. The
distinction matters because the two are recovered under different assumptions
and, as our transfer results show, survive domain change to different degrees.

Crucially, $y$ is the \emph{coded} label, not ground truth. Where an agency
fails to record animal involvement, $y_i = 0$ for a record that is in fact a
WVC. We put this asymmetry to work: a model fitted to partially corrupted labels
still assigns high scores to unrecorded WVCs, because their covariate signature
is unchanged.

\subsection{Label-Leakage Screen}

Crash schemas contain fields that encode the outcome directly. A
first-harmful-event field with the value \emph{animal} is not a predictor of a
WVC; it is the label under another name. Models built on such fields report
near-perfect accuracy that reflects nothing but circularity, a failure mode
documented across seventeen scientific fields by \citet{kapoor2023} under the
heading of feature leakage. No prior WVC study we are aware of screens for it.

We therefore apply an explicit screen before any model is fitted. A candidate
field $f$, taking value $X_f$ in $\mathcal{V}_f$, is excluded when some single value is both overwhelmingly associated with the positive class and accounts for a
substantial share of all positives:

\begin{equation}
\label{eq:leak}
\text{exclude } f
\quad\Longleftrightarrow\quad
\text{some value } v \text{ has}\quad
\underbrace{\Pr(y{=}1 \mid X_f{=}v)}_{\text{purity}} \ge \tau_r
\quad\text{and}\quad
\underbrace{\Pr(X_f{=}v \mid y{=}1)}_{\text{coverage}} \ge \tau_c .
\end{equation}
\hfill\break

\noindent Purity is the share of records carrying value $v$ that are WVCs;
coverage is the share of all WVCs that carry value $v$. A field is excluded when
some value scores high on both, with $\tau_r = 0.85$ and $\tau_c = 0.15$. These thresholds are discretionary: the appropriate value depends on the schema being screened, and the leakage literature offers no general criterion, only qualitative detection heuristics that require domain knowledge \citep{kaufman2012}. Both screen conditions are important. Purity alone would discard a rare value that happens to
be wildlife-dominated but explains almost nothing; coverage alone would discard
common values with no diagnostic power. The pair isolates fields that function
as a restatement of the label while leaving genuinely predictive covariates
intact. Single-vehicle involvement, for instance, carries
$\Pr(y=1 \mid \text{single vehicle}) \approx 0.19$ in Colorado and is retained.

The screen deliberately runs over a candidate set that \emph{includes} the
suspect fields, so that we can audit its decisions, and its output is reported as
a substantive result. Identifiers, names, coordinates,
free text, raw timestamps, and outcome counts are excluded beforehand as
non-descriptors. Three derived temporal fields (month, hour of day, and day of
week) are added for every state; they cannot encode the label and are available
universally.

\subsection{Redundancy Screen}

Leakage and collinearity are separate concerns, so we also screen for redundant
predictors before interpreting importances. Variance inflation factors and
Pearson correlations assume continuous, roughly linear predictors, whereas these
fields are almost all categorical and the few numeric ones are heavily
right-skewed. We therefore measure association with a mixed-type, distribution-free
matrix: bias-corrected Cram\'er's $V$ for two categorical fields, the correlation
ratio for a categorical field against a numeric one, and the absolute Spearman
coefficient for two numeric fields, each scaled to $[0,1]$ so a single benchmark
applies. Temporal fields count as categorical, and field typing follows the model
exactly, so the redundancy we measure is the redundancy the model sees. Any pair reaching $0.80$ is reduced by the greedy correlation-filter rule of \cite{kuhn2008}, which drops the field with the highest mean association until no surviving pair clears the benchmark. The adopted cutoff sits between the ecological convention of 0.7 \citep{Dormann2013} and the caret default of 0.90 \citep{kuhn2008}.

Across the four states only two pairs reach the benchmark, both in Illinois and both involving the national-highway-system indicator, which is $0.97$ associated with class of trafficway and $0.95$ with roadway functional class. The rule resolves the resulting three-field cluster by dropping that single indicator. Both decisions rest on Cram'er's $V$, which carries no distributional assumption. The
only skew-sensitive measure, the correlation ratio, arises for one field in
Colorado with a maximum association of 0.24; recomputing it on log and rank
transforms raises that value only to 0.30, well below the benchmark. No merge
therefore depends on the distributional form of the data.

\subsection{Classifier and Calibration}

Crash records are wide, categorical, and non-smooth in the target, which is the
regime in which tree ensembles retain a decisive advantage over deep
architectures even under heavy tuning \citep{grinsztajn2022}. We use CatBoost
\citep{prokhorenkova2018}, whose ordered target statistics handle the
high-cardinality categorical fields natively without the target leakage that
ordinary target encoding introduces, an algorithmic counterpart to the
data-level screen of Equation~\ref{eq:leak}. Models use 400--600 boosting iterations at
depth 8 and learning rate 0.08, with balanced class weights and a fixed seed.
Random forests \citep{breiman2001} and histogram gradient boosting
\citep{pedregosa2011} are fitted as controls to establish that results reflect
the data rather than the estimator.

Boosted trees rank well but return distorted probabilities, and ranking quality
and calibration quality are decoupled \citep{guo2017}. Prevalence estimation
consumes posteriors rather than ranks, so calibration is not optional here. We
fit an isotonic calibrator on a held-out calibration split, following the
demonstration by \citet{niculescumizil2005} that isotonic regression
specifically repairs boosted full trees. Training data are split 80/20 into
fitting and calibration partitions, stratified jointly by state and label, with
the test partition held out entirely. Throughout, we write
$p_i = \Pr(y{=}1 \mid \mathbf{x}_i)$ for the calibrated posterior of record $i$,
the quantity the estimators below consume.

\subsection{Two Feature Regimes}

Two questions require two feature spaces, and conflating them would confound the
results.

For per-state detection, richness analysis, and explanation, each state is
modelled on \emph{its own} leakage-screened fields. This measures what a state
can achieve with the record it actually keeps, which is the quantity a data
manager cares about, and it is the only regime in which the richness comparison
is meaningful.

For cross-state transfer and quantification, a common feature space is
unavoidable. We harmonize to fourteen variables shared by all four states: eight
numeric (single-vehicle indicator, annual average daily traffic, posted speed,
lane count, sinuosity, month, hour, day of week) and six categorical (weather,
lighting, surface condition, severity, junction type, urban--rural). Each
state's native vocabulary is mapped to a shared code list: weather to clear,
rain, snow, fog, or wind; lighting to daylight, dark-lighted, dark-unlighted, or
dusk-dawn; severity to property-damage-only, injury, or fatal. Junction type
required state-specific derivation to remain non-circular, since Iowa's obvious
junction fields fail the leakage screen; there it is derived from whether the
location description names two cross-streets. Numeric fields are median-imputed
and categorical fields mode-imputed and one-hot encoded.

\subsection{Evaluation}

With positives rare and their rate differing across states, average
precision is the primary metric. Under class imbalance the receiver operating
characteristic is dominated by true negatives and reports optimistically, while
the precision--recall baseline equals prevalence and therefore remains
informative \citep{saito2015}. Average precision is computed as the
prevalence-weighted mean of precision over the recall range,

\begin{equation}
\label{eq:ap}
\mathrm{AP} = \sum_{k} \left( R_k - R_{k-1} \right) P_k ,
\end{equation}
\hfill\break
\noindent with $P_k$ and $R_k$ the precision and recall at the $k$th threshold.
Average precision approaches $\pi$ for an uninformative model and equals one for a
perfect ranking, so raw values are not comparable across states of differing
prevalence. Alongside each value we therefore report a prevalence-neutral skill
score that rescales average precision to the interval between chance and
perfection,

\begin{equation}
\label{eq:skill}
S = \frac{\mathrm{AP} - \pi}{1 - \pi} ,
\end{equation}
\hfill\break
\noindent where $S = 0$ corresponds to chance performance and $S = 1$ to a
perfect ranking.
Precision, recall, and $F_1$ are reported at the threshold maximizing $F_1$ on
the precision--recall curve. All headline metrics carry 95 percent percentile
confidence intervals from 500 test-set bootstrap resamples.

Random splits would overstate transferability, since state boundaries coincide
with reporting regime, road network, and species assemblage. Following
\citet{roberts2017}, each state is treated as a block and transfer is evaluated
leave-one-state-out.

\subsection{Isolating Richness from Rarity}

A four-state comparison confounds record richness with prevalence, labelling
convention, and reporting culture. To separate them we vary richness
\emph{within} each state, holding everything else fixed. Fields are ranked by
importance from a full-feature fit, then re-fitted cumulatively, adding one
field at a time and recording average precision with 95 percent bootstrap
intervals at each step. Training and test partitions are capped at 400,000 and
150,000 records to keep the $\sum_s |\mathcal{F}_s| = 61$ fits tractable,
where $\mathcal{F}_s$ is state $s$'s field set. A
rising curve within a single state, at a single prevalence, is direct evidence
that the recorded field set governs detection. Comparing curve endpoints against
a matched field budget then separates the state effect from the richness effect.

\subsection{Prevalence Estimation Without Labels}

The audit contribution rests on recovering $\pi$ in a target sample whose labels
are unavailable or untrustworthy. We compare three estimators. Classify-and-count
simply counts positive decisions,

\begin{equation}
\label{eq:cc}
\hat{\pi}_{\mathrm{CC}} = \frac{1}{n}\sum_{i=1}^{n} \mathds{1}\!\left[\, p_i \ge t \,\right] ,
\end{equation}
\hfill\break
\noindent the fraction of records whose posterior $p_i$ clears the decision
threshold $t$, and is biased whenever target prevalence departs from training
prevalence \citep{forman2008}. Adjusted classify-and-count inverts the
classifier's true- and false-positive rates ($\mathrm{TPR}$, $\mathrm{FPR}$), estimated on the calibration partition, and clips the
result to the unit interval:

\begin{equation}
\label{eq:acc}
\hat{\pi}_{\mathrm{ACC}} = \mathrm{clip}_{[0,1]}\!\left(
\frac{\hat{\pi}_{\mathrm{CC}} - \mathrm{FPR}}{\mathrm{TPR} - \mathrm{FPR}} \right) .
\end{equation}
\hfill\break
\noindent Equation~\ref{eq:acc} is unbiased when the class-conditional score
distributions are stable between source and target, and becomes unstable as the
denominator shrinks, the failure mode \citet{forman2008} identifies under
severe imbalance.

Maximum-likelihood prevalence estimation makes no use of a threshold. Starting
from a model trained at source prevalence $\pi_s$, it re-estimates the target
prevalence by expectation-maximization \citep{saerens2002}, reweighting each
posterior $p_i$ toward the current prevalence estimate and averaging. Writing $\pi^{(m)}$ for the estimate at iteration $m$, the update is:

\begin{equation}
\label{eq:mlpe}
\pi^{(m+1)} = \frac{1}{n} \sum_{i=1}^{n}
\frac{(\pi^{(m)}/\pi_s)\, p_i}
{(\pi^{(m)}/\pi_s)\, p_i + \bigl((1-\pi^{(m)})/(1-\pi_s)\bigr)(1 - p_i)} .
\end{equation}
\hfill\break
\noindent The iteration starts from $\pi^{(0)} = \tfrac{1}{n}\sum_i p_i$ and
runs until the estimate changes by less than $10^{-7}$. The iteration maximizes the likelihood of the
unlabelled target sample and returns the adjusted prior as its output. Consuming
posteriors rather than decisions, it inherits the calibration dependence that
\citet{esuli2021} identify as its principal weakness, which is why the
calibration step of the preceding subsection is load-bearing rather than
cosmetic.

Following the evaluation protocol of \citet{gonzalez2018}, accuracy is assessed
by systematically varying target prevalence rather than at a single operating
point. The held-out test set is resampled to 120,000 records at true prevalences
from 2 to 20 percent in steps of two, and each estimator is scored by absolute
error. The same three estimators are then applied across states, training on
three and estimating the fourth, where the shift is not controlled but observed.

\subsection{Under-Reporting Audit}

Equation~\ref{eq:mlpe} reads prevalence from covariate signatures rather than
from the animal flag, so its estimate is insensitive to whether the agency coded
a given crash correctly. The under-reporting estimate follows directly as

\begin{equation}
\label{eq:ur}
\widehat{\Delta} = \hat{\pi}_{\mathrm{MLPE}} - \pi_{\mathrm{recorded}} ,
\end{equation}

\noindent the excess of estimated over coded prevalence. Validating
$\widehat{\Delta}$ requires a ground truth that no real crash file provides, so
we inject one: a known fraction $\phi \in \{0, 0.25, 0.50, 0.75\}$ of confirmed
wildlife crashes is relabelled as non-wildlife, driving the recorded rate to
$\pi(1-\phi)$ while the true rate is unchanged, and $\widehat{\Delta}$ is
compared against the known gap $\pi\phi$.

A complementary experiment tests the record-level counterpart. Rather than
correcting an aggregate, we ask whether the individual hidden crashes can be
surfaced: the model is retrained on the corrupted labels, and we measure what
fraction of the hidden WVCs it still flags at the operating point an agency
would have chosen, together with where those crashes sit in the score
distribution of records coded non-animal.

\section{RESULTS}

\subsection{The Leakage Screen and Iowa's Under-Coding}

Equation~\ref{eq:leak} screens the data before any model is fitted, and it
removed the fields that restate the outcome. The resulting pattern is itself informative
(Table~\ref{tab:leakage}). Colorado, Utah, and Illinois each lose only the
obvious offenders, a first-harmful-event or crash-type field that is all but
synonymous with the animal flag. Iowa loses ten of fifteen. Past the expected
harmful-event field, a single sentinel value, 77, recurs across weather,
lighting, surface condition, road type, and manner of collision, each roughly 99
percent wildlife and each covering three-quarters of Iowa's WVCs. That value
falls outside the real code range for those fields, so it marks not-applicable:
Iowa routes animal crashes through an abbreviated report that leaves the ordinary
descriptors blank. The record thins exactly where the crash is a WVC, which
leaves Iowa with only eight usable fields and, as the detection results show
next, the weakest prevalence-adjusted performance of the four. Any agency can
apply the same screen to its own file to identify where its coding practice breaks
down.

\begin{table}[!ht]
\centering
\caption{Fields removed by the leakage screen.}
\label{tab:leakage}
\begin{tabular}{|l|l|r|r|}
\hline
\textbf{State} & \textbf{Flagged field} & \textbf{Purity (\%)} & \textbf{Coverage (\%)} \\
\hline
\multirow{3}{4em}{Colorado} & first harmful event            & 100 & 87.8 \\
%\hline
         & most harmful event             & 100 & 86.2 \\
%\hline
         & crash type                     & 100 & 89.8 \\
\hline
\multirow{1}{4em}{Utah} & first harmful event            &  99.9 & 98.3 \\
\hline
\multirow{10}{4em}{Iowa}     & first harmful event            & 100 & 100 \\
%\hline
         & major cause                    &  99.3 & 100 \\
%\hline
         & manner of collision            &  99.6 & 42.5 \\
%\hline
         & weather                        &  98.7 & 78.5 \\
%\hline
         & surface condition              &  98.7 & 78.5 \\
%\hline
         & lighting                       &  98.7 & 77.6 \\
%\hline
         & road type                      &  98.6 & 72.5 \\
%\hline
         & location of first harm         &  97.8 & 56.9 \\
%\hline
         & roadway contributing circ.     &  93.7 & 75.8 \\
%\hline
         & environ. contributing circ.    &  92.5 & 46.4 \\
\hline
\multirow{2}{4em}{Illinois} & type of first crash            & 100 & 100 \\
%\hline
         & primary cause                  &  94.8 & 92.1 \\
\hline
\end{tabular}

\vspace{0.4em}
\begin{minipage}{\linewidth}
\footnotesize A field is flagged when, for some value $v$, purity
$\Pr(y{=}1 \mid X_f{=}v) \ge 85\%$ and coverage $\Pr(X_f{=}v \mid y{=}1) \ge 15\%$
(Equation~\ref{eq:leak}). Both columns report the same flagged value~$v$: the
field value most strongly associated with wildlife crashes.
\end{minipage}
\end{table}

\subsection{Across States: The Record, Not Prevalence, Sets Detection}

\begin{figure}[!ht]
  \includegraphics[width=\linewidth]{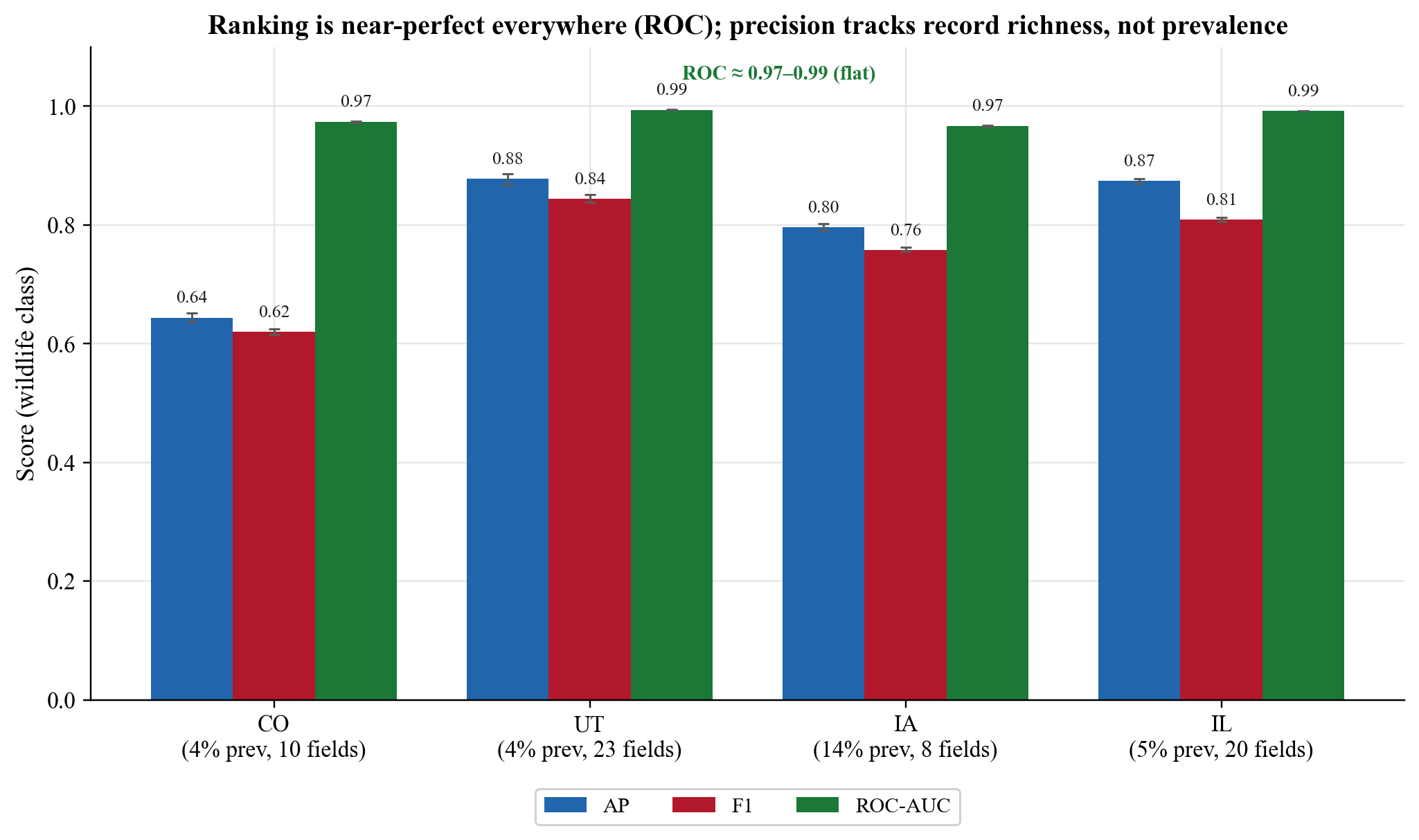}
  \caption{Ranking is near-perfect everywhere; average precision tracks record
  richness, not prevalence.}
  \label{fig:detection}
\end{figure}

With the circular fields gone, every state model still separates wildlife crashes
from the rest of its file well above chance. Ranking is near-perfect everywhere,
with ROC-AUC from 0.967 in Iowa to 0.994 in Utah (Table~\ref{tab:perstate}).
However, average precision better establishes the core argument of this paper.
Utah reaches 0.878 and Illinois 0.874, Iowa attains 0.796, and Colorado trails at
0.644 (Figure~\ref{fig:detection}).

The most informative comparison is Utah against Colorado. The two states carry
almost identical WVC prevalence, 4.4 percent and 3.8 percent, so prevalence is
held roughly constant between them, yet Utah's average precision exceeds
Colorado's by a quarter of the scale. The two states differ in the record: Utah
describes a crash with 23 screened fields and Colorado with 10. Prevalence cannot
account for the difference, whereas richness can. The prevalence-neutral skill
score of Equation~\ref{eq:skill} confirms the same ordering across all four
states. The two feature-rich states rank highest and nearly equal, Utah and
Illinois at 0.87, feature-poor Colorado ranks lowest at 0.63, and Iowa attains
only 0.76 despite carrying the highest prevalence of the four. Detection
therefore follows the completeness of the record rather than the rarity of the
event, which reverses the emphasis the imbalanced-learning literature places on
prevalence as the binding constraint.

\begin{table}[!ht]
\centering
\caption{Per-state in-domain detection, with 95\% bootstrap intervals.}
\label{tab:perstate}
\begin{tabular}{|l|r|r|c|c|r|r|}
\hline
\textbf{State} & \textbf{Prev.} & \textbf{Fields} & \textbf{AP [95\% CI]} & \textbf{ROC [95\% CI]} & \textbf{$F_1$} & \textbf{Skill} \\
\hline
Colorado & 3.75\%  & 10 & 0.644 [0.636, 0.652] & 0.974 [0.973, 0.975] & 0.620 & 0.63 \\
\hline
Utah     & 4.38\%  & 23 & 0.878 [0.868, 0.886] & 0.994 [0.993, 0.995] & 0.844 & 0.87 \\
\hline
Iowa     & 13.96\% &  8 & 0.796 [0.790, 0.802] & 0.967 [0.966, 0.968] & 0.758 & 0.76 \\
\hline
Illinois & 5.30\%  & 20 & 0.874 [0.870, 0.878] & 0.992 [0.992, 0.992] & 0.809 & 0.87 \\
\hline
\end{tabular}

\vspace{0.4em}
\begin{minipage}{\linewidth}
\footnotesize AP, average precision; Skill, the prevalence-neutral score
$(\mathrm{AP} - \pi)/(1 - \pi)$ of Equation~\ref{eq:skill}, where $0$ is chance
and $1$ is perfect; $F_1$ at the $F_1$-optimal threshold. Intervals from 500
test-set bootstrap resamples.
\end{minipage}
\end{table}

\subsection{Within a State: Isolating Richness from Prevalence}

\begin{figure}[!ht]
  \includegraphics[width=\linewidth]{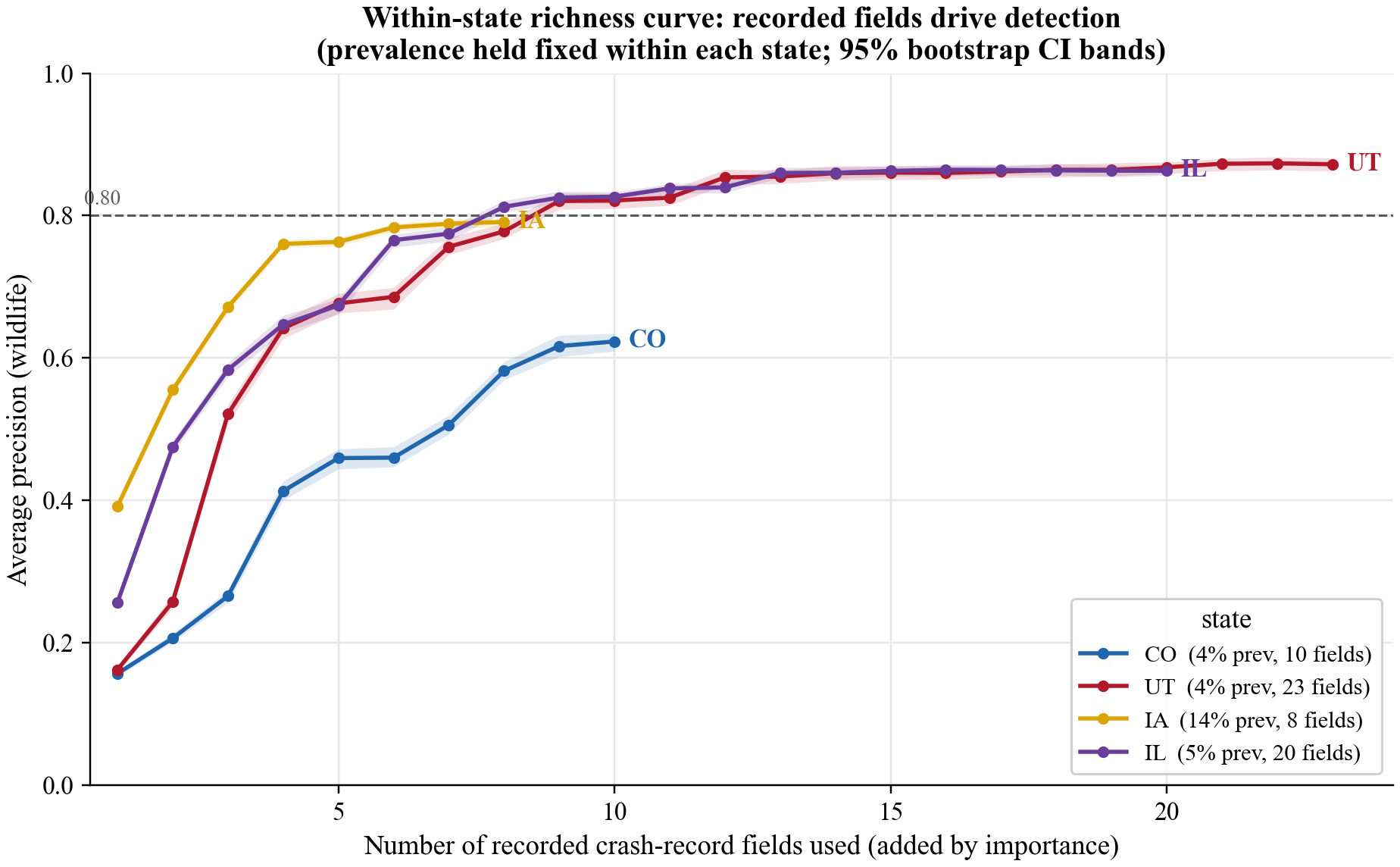}
  \caption{Within-state richness curves: average precision rises as fields are
  added, prevalence fixed.}
  \label{fig:richness}
\end{figure}

The cross-state comparison entangles richness with everything else that separates
one state from another. The within-state curve removes that confound by adding a
state's own fields one at a time, prevalence fixed throughout
(Figure~\ref{fig:richness}).
Detection climbs with every field in all four states. Utah rises from an average
precision of 0.16 on its single strongest field to 0.87 on all 23; Colorado
climbs the same way but exhausts its record at 10 fields and plateaus at 0.62.
Held to a matched eight-field budget, Utah (0.78) and Iowa (0.79) finish level
despite Iowa carrying three times Utah's prevalence, and Colorado lags at 0.58.
Same state, same prevalence, more of the record, and detection improves. This is
the direct evidence that record completeness, not event prevalence, governs how
well WVCs can be found.

\subsection{What the Models Read}

\begin{figure}[!ht]
  \includegraphics[width=\linewidth]{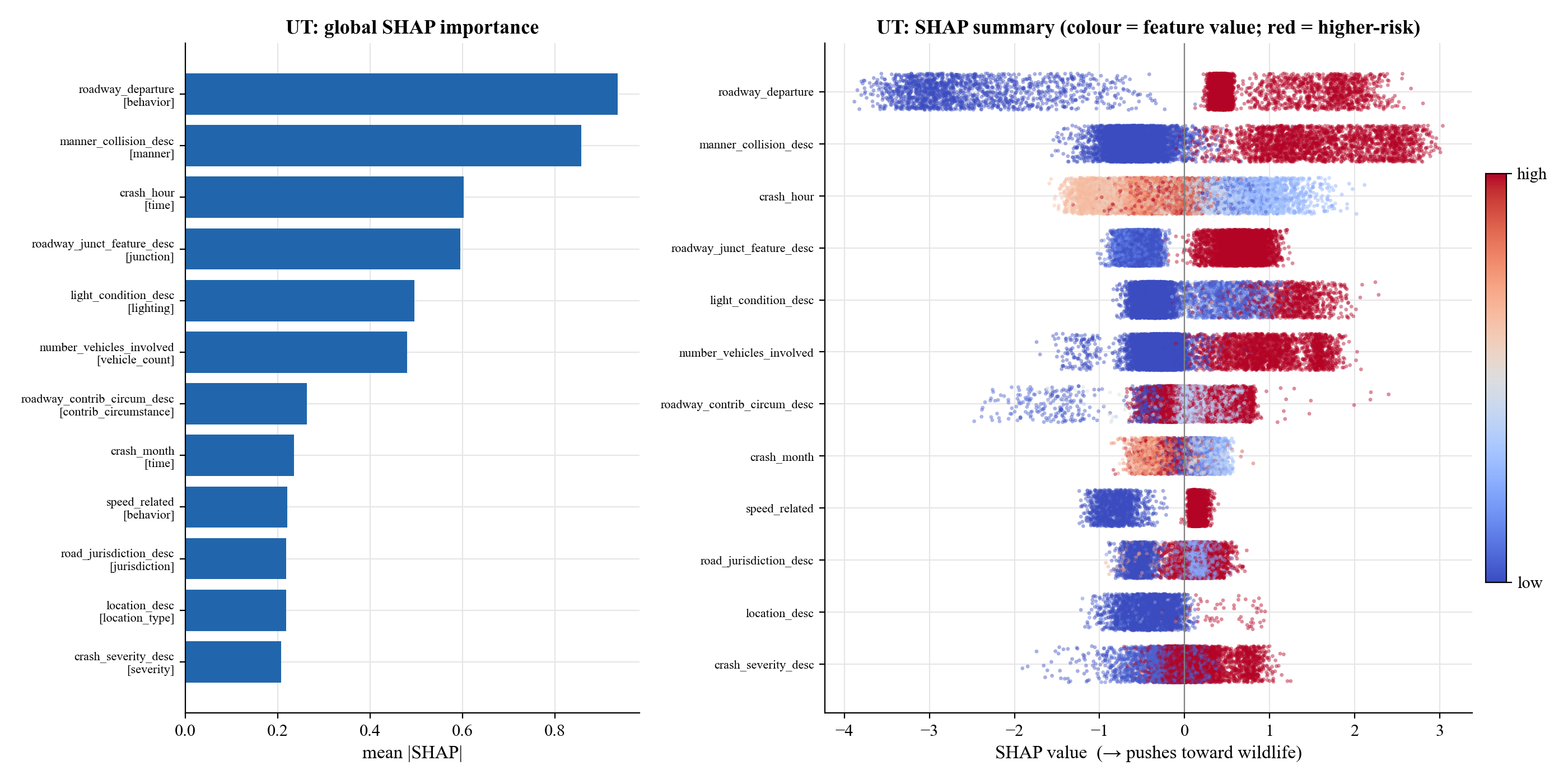}
  \caption{Utah: global Shapley importance (left) and per-record effect
  direction (right).}
  \label{fig:shap}
\end{figure}

Reading Utah, the richest record, with Shapley values shows a coherent physical
signature (Figure~\ref{fig:shap}). The
strongest contributors are roadway departure, manner of collision, hour of day,
junction feature, lighting, and vehicle count, and their directions align:
single-vehicle, run-off-road crashes on open non-intersection segments, in
darkness, away from peak hours, push a record toward wildlife. That is the
profile of an animal entering a rural road at night.

Aggregating importance by concept across all four states yields a concrete
data-collection recommendation. Time of day and season lead (present in all four states,
mean importance share 0.26), followed by vehicle count (all four, 0.22), junction
type (three states, 0.10), severity (all four, 0.06), and lighting (three states,
0.05). None of these is exotic or expensive to record. Five fields most agencies
already capture carry the bulk of the detectable signal, which is what makes the
detection result portable to states beyond these four.

\subsection{What Transfers Across States}

A model trained on three states and applied to the fourth keeps its ranking
almost intact: ROC-AUC on the held-out state stays between 0.964 and 0.974,
close to the in-domain values of Table~\ref{tab:perstate}. Decisions are another
matter. Setting the held-out threshold from prevalence alone recovers 97 to 98
percent of the oracle $F_1$ for Colorado and Utah and 82 percent for Iowa, then
collapses to zero for Illinois. The Illinois failure has a precise explanation:
the transferred model's scores on Illinois are compressed so low that
the $F_1$-optimal threshold is 0.017, while the prevalence rule prescribes 0.41,
so the model flags almost nothing. Ranking survives the border; calibration does
not. Cross-state, a borrowed model is an instrument for screening and ranking,
not for per-record decisions at a transplanted threshold.

\subsection{Prevalence, the Under-Count, and Recovering Hidden Records}

\begin{figure}[!ht]
  \includegraphics[width=\linewidth]{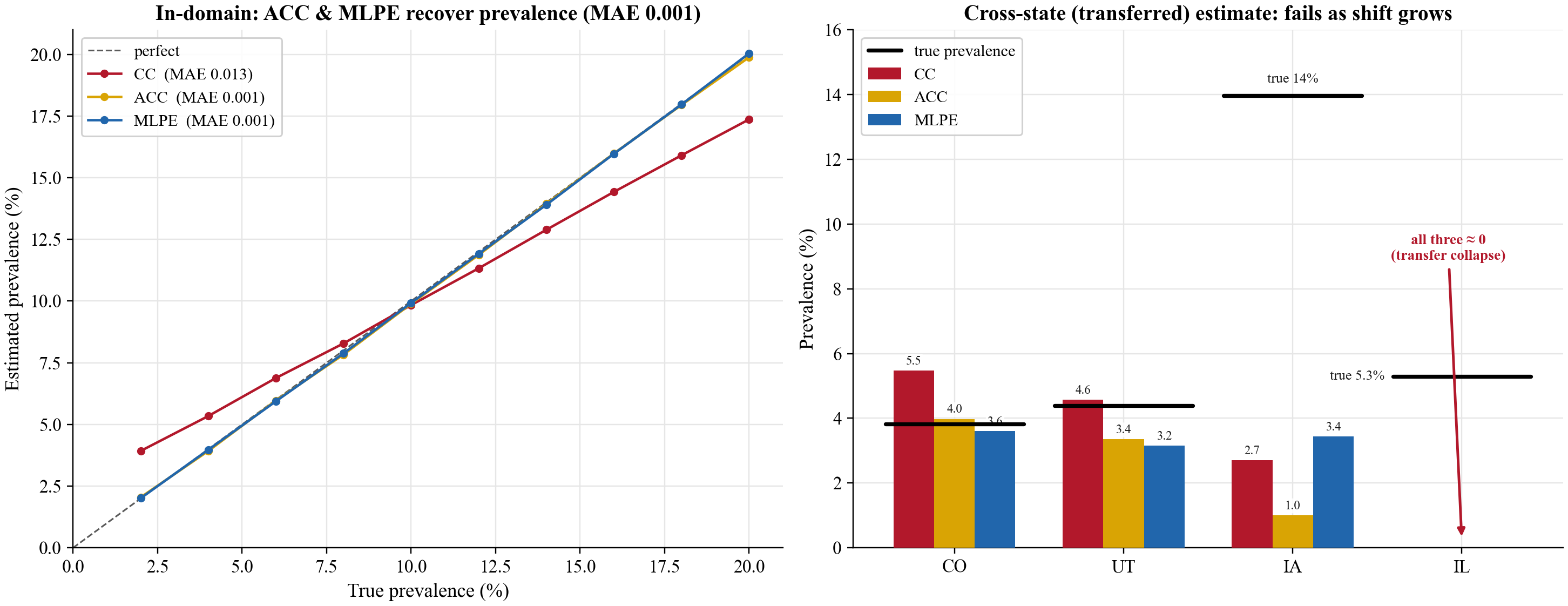}
  \caption{Prevalence estimated from scores: in-domain against known prevalence
  (left) and across states (right).}
  \label{fig:quant}
\end{figure}

Read from the scores, prevalence recovers with striking accuracy in-domain,
independent of the coded flag (Figure~\ref{fig:quant}). Across injected
prevalences from 2 to 20 percent, mean absolute error is 0.0006 for
maximum-likelihood estimation and 0.0008 for adjusted classify-and-count,
against 0.0127 for naive classify-and-count, whose error grows as prevalence
drifts from the training rate. Across states the estimators behave exactly as the
transfer analysis predicts. Where score distributions align they track truth,
estimating Colorado's 3.8 percent at 3.6 and Utah's 4.4 percent within a point.
Where the shift is severe they fail together, reading Iowa's 14 percent as 1 to 3
and Illinois's 5.3 percent as effectively zero, the same Illinois collapse the
threshold transfer exposed. Prevalence estimation inherits the ranking that
crosses borders and the calibration that does not.

The under-count audit follows from this in-domain accuracy. Injecting a known
reporting gap into Colorado drives the recorded rate down while the score-based
estimate holds at the true 3.8 percent, since the scores do not depend on whether
a crash was coded correctly. The recovered gap matches the injected gap almost
exactly: 1.89 percent estimated against 1.91 true with half of WVCs uncoded, and
2.84 against 2.86 with three-quarters uncoded. On its own crash file, an agency
can quantify the extent to which it under-reports its wildlife collisions.

The audit reaches individual records as well as the aggregate. Corrupting the
labels to simulate under-reporting, by hiding a fraction of known wildlife
crashes and retraining on the result, fails to hide those crashes from the model.
Whatever the fraction concealed, the hidden WVCs land at the 97th to 99th
percentile of records coded non-animal, and with half of them hidden the model
re-flags 86 percent at the operating point an agency would use. The same model
reconstructs descriptors left blank, predicting each held-out field from the
others well above its majority baseline: surface condition at 0.92 against 0.82,
lighting at 0.84 against 0.70, weather at 0.81 against 0.71, and junction at 0.63
against 0.47. One model detects the collision, surfaces the crashes an agency
failed to log, and backfills the fields it left blank.

\section{DISCUSSION}

The prevailing approach to wildlife crashes in imbalanced data treats their
scarcity as the central obstacle and applies resampling to address it. Our
results locate the constraint elsewhere. Utah and Colorado detect at nearly the
same prevalence, yet Utah's average precision exceeds Colorado's by roughly a
quarter of the scale because Utah records more than twice as many fields. The
within-state curves demonstrate this directly: with a state held fixed and its
own fields added one at a time, detection improves at each step. Detection is
limited by the completeness of the report, and no synthetic rebalancing recovers
information a form never captured. The record itself is therefore where an agency
can intervene.

This reframing changes what a department can do with records it already holds. A
calibrated model trained on screened fields surfaces wildlife crashes an officer
never coded as such. With half of Colorado's WVCs concealed, the hidden crashes
still sit at the 98th percentile of records marked non-animal, and the model
re-flags most of them at an ordinary operating point. The same model imputes
descriptors left blank, recovering surface condition and lighting well above
their base rates. Run over a legacy file, it becomes an instrument for
retrospective cleanup and quality control. The leakage screen carries its own
diagnostic, and applied to Iowa it exposed a coding path that stamps a
not-applicable sentinel across five ordinary fields on animal crashes, thinning
the record exactly where the crash is a WVC. Any department can apply the same
screen to its own data to identify where its reporting practice breaks down.

The prevalence machinery turns the classifier into an auditing instrument.
In-domain it recovers the true WVC rate to within 0.001 across a tenfold range of
prevalence, and it reconstructs an injected reporting gap almost exactly,
estimating a 1.9 percent under-count against a 1.9 percent truth. An agency can
therefore put a defensible number on how many wildlife collisions its file omits,
working from that file alone. That number converts to dollars through established
per-collision costs and gives mitigation spending an evidence base, and it needs
no carcass-removal survey to produce, which is the second data source the closest
prior method requires \citep{zou2019}.

Cross-state behaviour marks the boundary of the method honestly. Ranking survives
a state border almost intact, with held-out ROC-AUC between 0.96 and 0.97, so a
borrowed model still separates likely wildlife crashes from the rest. Calibration
does not survive, and decisions set at a transplanted threshold can collapse, as
they did for Illinois. A model trained elsewhere therefore serves for screening
and for prevalence estimation where score distributions align, and it should carry
local calibration before it renders per-record verdicts in a new state. The
concept importances point to a short, actionable list for reporting manuals. Time
of day and season, vehicle count, junction type, severity, and lighting carry
most of the detectable signal, and all five are inexpensive and already common.
States that record them well make every result here reproducible on their own
data.

Several assumptions bound these claims. Four states support the richness argument
within each of them, though a wider sample would test how far it generalizes. The
animal label differs across states, wild-only in Colorado and Utah and broader in
Iowa and Illinois, so the per-state results stand on their own footing while
cross-state prevalence comparisons carry that caveat. The under-count audit
assumes reporting failures fall roughly evenly across the recorded fields. Where
an agency omits a distinct slice of crashes, the covariate signature of the
missing records shifts, and recovery degrades through the same mechanism that
collapses cross-state estimation under severe distribution shift. The
under-reporting validation also injects a known gap into a single state, so it
demonstrates the mechanism rather than a field-measured rate. Confirming the audit
against independent carcass counts is the natural next step.

\section{CONCLUSION}

We recast wildlife-vehicle collision surveillance from a spatial question, where
collisions concentrate, to a record question, which crashes are collisions and
how many go uncounted. Across 4.9 million
crashes in four states, calibrated gradient-boosted models detect wildlife
crashes from ordinary fields alone, and the completeness of the record, not the
rarity of the event, sets how well they detect. Read as scores in place of
labels, the same models estimate a state's true WVC prevalence and measure the
gap between that estimate and the rate the agency recorded. The returns for
practice are immediate: surfacing collisions never logged, backfilling missing
fields, quantifying a true collision burden for mitigation funding, and naming
the handful of fields a reporting manual should require. Extending the framework
across more states, and validating the under-count against independent carcass
data, would carry a demonstrated instrument into standard transportation
practice.

\section{ACKNOWLEDGMENT}
The authors acknowledge the use of Anthropic (2026), Opus 4.8 for controlled python code generation, code audit, compilation of additional relevant sources, grammar and writing audits only. 

\section{AUTHOR CONTRIBUTIONS}
The authors confirm contribution to the paper as follows: study conception and design: \textbf{\textit{R. Bello, Varghese Vaidyan, J. Mwakalonge}}; data collection: \textbf{\textit{R. Bello, J. Mwakalonge}}; analysis and interpretation of results: \textbf{\textit{R. Bello, G. Comert, V. Bendigeri}}; draft manuscript preparation: \textit{\textbf{R. Bello, A. Dontoh, M. Mkufya}}. All authors reviewed the results and approved the final version of the manuscript.

\section{DECLARATION OF CONFLICTING INTERESTS}
The authors declared no potential conflicts of interest with respect to the research, authorship, and/or publication of this article.

\section{FUNDING}
This research was partly funded by the U.S. Department of Education through the HBCU Master’s Program Grant (Grant No. P120A210048); the U.S. Department of Transportation’s University Transportation Centers Program grant, administered by the Transportation Program at South Carolina State University (SCSU); and the National Science Foundation (NSF) under Grant Nos. 2131080, 2242812, 2200457, 2234920, and 2305470.

\clearpage
\nolinenumbers
\printbibliography[title={REFERENCES}]

\end{document}